\documentstyle[aps,twocolumn]{revtex}

\begin{document}
\draft 
\preprint{\today}
\title{
Numerical study of disorder effects
on the three-dimensional Hubbard model
}
\author{Y. Otsuka$^1$ and Y. Hatsugai$^{1,2}$}

\address{
$^1$Department of Applied Physics, University of Tokyo,
7-3-1 Hongo Bunkyo-ku, Tokyo 113-8656, Japan\\
$^2$PRESTO, Japan Science and Technology Corporation, Japan
}
\maketitle

\begin{abstract}
Combined effects of interactions and disorder
are investigated using
a finite temperature quantum Monte Carlo technique 
for the three-dimensional Hubbard model 
with random potentials of a finite range.
Temperature dependence of the charge compressibility 
shows that the Mott gap collapses 
beyond a finite disorder strength.
This is a quantum phase transition 
from an incompressible phase
to a compressible phase driven by disorder.
We calculate the antiferromagnetic structure factor
in the presence of disorder as well.
Strong antiferromagnetic correlation,
which is characteristic of the Mott insulator,
is destroyed by a finite amount of disorder.
\end{abstract}


\section{Introduction}
As well as the repulsive interaction between electrons,
the influence of disorder is essential 
in many electronic systems.
Although either of these two effects 
causes a metal-insulator transition,
the physical characters are quite different.
The repulsive interaction tends to 
suppress the double occupancy of electrons.
On the other hand, in random potentials,  
electrons can favor doubly occupied states
if the random potential is sufficiently low at the site.
Therefore, the interaction and disorder 
may have opposite effects 
on the charge degree of freedom.
In the insulator due to the interaction (Mott insulator),
the charge fluctuation is strongly suppressed and 
a finite charge gap opens,
while the insulating phase due to disorder (Anderson insulator) 
does not necessarily have a charge gap.
Another difference between these two insulators is
the existence of magnetic correlation.
Since the repulsive interaction
induces local magnetic moments,
the Mott insulator has a strong antiferromagnetic correlation.
On the other hand,
the Anderson insulator does not necessarily
enhance magnetic correlation.
Therefore one may expect that the interaction and disorder compete
in both charge and spin degree of freedom. 
Especially, it is important to investigate 
to what extent the stability of the Mott insulator remains 
in the presence of disorder.

The Hubbard model with disorder 
is one of the simplest models that include these two effects.
In one and two dimensions,
the transition from the Mott to the Anderson insulator is confirmed
by various methods \cite{ref-Suzumura} - \cite{ref-Kishine}.
Also the dynamical mean field theory \cite{ref-dmft} is applied
to the infinite dimensional Hubbard model and consistent results are obtained
\cite{ref-Ulmke}.
On the other hand, there is no work beyond the mean field approximation
in three-dimensional case \cite{ref-Tusch}.
Therefore approximation-free results can be useful to understand 
the three-dimensional strongly correlated system with disorder.

In the present paper, we study 
the three-dimensional Hubbard model with random potentials 
using a finite temperature quantum Monte Carlo (QMC) method.
The rest of this paper is organized as follows.
In Sec.II,
we introduce the three-dimensional disordered Hubbard model 
and describe physical observables.
In Sec.III, 
we discuss the effects of disorder on the charge compressibility and 
the magnetic structure factor.

\section{Model and Method}
The Hamiltonian of the disordered Hubbard model is given by
\begin{equation}
  \label{hamiltonian}
{\cal H}=
-t \sum _{\langle i,j \rangle  \sigma}
(\hat{c}_{i \sigma }^{\dagger}\hat{c}_{j \sigma} + 
 \hat{c}_{j \sigma}^{\dagger}\hat{c}_{i \sigma})
+ U \sum _{i}\hat{n}_{i \uparrow}\hat{n}_{i \downarrow}
+ \sum _{i \sigma} w_{i} \hat{n}_{i \sigma},
\end{equation}
where $t$ is the nearest neighbor hopping amplitude,
$\langle i,j \rangle$ is a nearest-neighbor link, 
$U$ is the on-site interaction
and 
$\{w_{i} \}$ are random potentials
chosen from a flat distribution 
in the interval [${\it -W}$,${\it W}$].
The system is on the cubic lattice in three dimensions
and we use the periodic boundary condition.
We treat the system in a grand canonical ensemble
with the chemical potential $\mu$. 
The grand canonical method is suitable to study 
the charge degree of freedom because
the charge fluctuation is 
taken into account statistically \cite{ref-Shiba}.
In this treatment, we make the system half-filled
by setting $\mu = U/2$.
In the absence of disorder 
with sufficiently large $U$, 
the ground state 
is an antiferromagnetic insulator 
where the charge fluctuation is strongly suppressed.

In order to obtain approximation-free results, we employ 
a finite temperature determinantal Quantum Monte Carlo method 
\cite{ref-afqmc1},\cite{ref-afqmc2}.
We also use the matrix-decomposition technique 
to remove numerical instabilities at low temperatures
\cite{ref-white}.
The simulations are performed 
in the half-filled sector ($\mu = U/2$) 
for lattices with sizes up to $N = 6 \times 6 \times 6 $ 
with $U/t=6$.
We choose a Trotter time slice size $\Delta\tau \simeq 0.15/t$.
We have checked that the systematic error due to 
the Suzuki-Trotter decomposition
is almost independent of temperatures and
does not change the qualitative feature.
For each realization of disorder, 
we have typically run 2000 Monte Carlo sweeps for measurements 
after 500 sweeps in the warming up run.
For all the observables, we average over 
24 realizations of disorder
and the errors are estimated by the variance among 
realizations of disorder.
Since 
the system does not have a particle-hole symmetry
in each realization of disorder,
the negative-sign problem occurs.
For example, the value of average sign is $\sim 0.1$ 
for $N = 4 \times 4 \times 4 $ and $W/t=1$
at temperature $T/t=0.1$.
Although it is not so severe as a doped case,
a simulation at a very low temperature with strong disorder
is difficult.

The physical observables we have calculated are the compressibility 
$\kappa$ and the magnetic structure factor $S(q)$ 
defined as
\begin{equation}
  \label{hub-kappa}
  \kappa = \frac{1}{N} \frac{\partial N_{e}}{\partial \mu} 
  = \frac{\beta}{N} 
\left( 
\langle \hat{N}_{e}^{2} \rangle - \langle \hat{N}_{e} \rangle ^{2}
\right) ,
\end{equation}
\begin{equation}
  \label{hub-sq}
  S(q) = \frac{1}{N} \sum_{i,j} e^{iq(r_{i} - r_{j})} 
\langle 
(\hat{n}_{i\uparrow}-\hat{n}_{i\downarrow})
(\hat{n}_{j\uparrow}-\hat{n}_{j\downarrow})
\rangle ,
\end{equation}
where $N$ is the number of sites,
$N_{e}$ is the number of electrons
and $\beta$ is an inverse temperature.
The charge compressibility $\kappa$
measures the charge fluctuation.
If the system has a finite charge gap, 
the compressibility shows thermally-activated behavior
in a low temperature region
and vanishes at $T=0$.
On the other hand, the system without a charge gap
has a finite compressibility at $T=0$
due to the existence of low-lying excitations.
The magnetic structure factor $S(q)$ at $q = (\pi,\pi,\pi)$ 
diverges in a low temperature region
when the system has an antiferromagnetic (quasi-) long-range order.

\section{Results and Discussion}
\label{sec:hub-results}

Figure 1 shows the temperature dependence of the charge compressibility
for several different strength of disorder.
Average over 24 realizations of disorder is performed.
Without disorder, 
the temperature dependence of the compressibility $\kappa$ shows 
thermally-activated behavior reflecting the existence of a finite charge gap.
The compressibility $\kappa$ at $T=0$ is zero 
within the numerical accuracy for the pure system.
It indicates that the ground state of the pure Hubbard model 
in three dimension is in an incompressible phase.
In the presence of disorder, 
the compressibility is enhanced.
For weak disorder, although enhanced, 
the compressibility still shows thermally-activated behavior
and the value at $T=0$ seems to be zero.
On the other hand, 
there is no tendency to decrease in $\kappa$ 
for $W > W_{c} (W_{c} \sim U/2)$ 
down to the lowest temperature we studied.
It means that the critical disorder strength to destroy
the Mott gap, $W_{c}$, is of the order of the Mott gap
since the system is in the strong coupling region ($U/t=6$).
Although we cannot exclude the possibility of
vanishing $\kappa$ at $T=0$,
what we have shown here is the best data
within the numerical restriction.
The results imply that
sufficiently strong disorder destroys
the Mott gap which is of the order of the interaction in the strong coupling region.
In the presence of disorder,
when we discuss the physics locally,
making one doubly occupied site gains
an potential energy $2W$ at the maximum,
while it costs a Coulomb energy $U$.
Therefore one may expect that 
the Mott gap collapses at $W > W_{c} (W_{c} \sim U/2)$ 
regardless of dimensionality
in the strong coupling region.
In other words, 
since 
charge properties of the Mott insulators
in the strong coupling region is determined locally,
the effect of disorder would be also local
and independent of dimensionality.
Indeed, the Mott gap collapses at $W_{c} \sim U/2$ 
in one\cite{ref-RAR}, two and infinite\cite{ref-Singh} dimensions
in the strong coupling region.
It is in contrast to the quantum nature of
the long-range properties of the correlation functions
which have a drastic difference in the dimensionality.
({\it e.g.} Luttinger liquid in one dimension).
The transition we observed is
a disorder-driven quantum phase transition
from an incompressible (gapped)
to a compressible (gapless) phase.
However, it does not necessarily mean 
an insulator-metal transition.
The compressibility takes a finite value in both a metallic phase
and an insulating phase due to disorder (Anderson insulator).
It is possible that 
the competition between the interaction and disorder
leads to a metallic phase especially in the three-dimensional system.
However, to distinguish these two phase, 
one needs simulations for a sufficiently large system, 
which we cannot perform because of the negative-sign problem.

Figure 2 shows the temperature dependence of 
the antiferromagnetic structure factor $S(\pi,\pi,\pi)$.
Since the ground state has an antiferromagnetic long-range order, 
the $S(\pi,\pi,\pi)$ shows diverging behavior 
toward the N\'{e}el temperature
in the absence of disorder.
For weak disorder,
the structure factor is slightly suppressed,
but diverging behavior is still observed
down to the temperature we studied.
This means that the ground state still has 
an antiferromagnetic long-range order.
When sufficiently strong disorder is included,
the temperature dependence of the $S(\pi,\pi,\pi)$
changes qualitatively.
The diverging behavior of $S(\pi,\pi,\pi)$ disappears.
This indicates that the long-range antiferromagnetic correlation
is also destroyed by a finite amount of disorder.
Ulmke {\it et al.} argue that 
weak disorder stabilizes antiferromagnetic order
for $U >U_{c}$, where $U_{c}$ is the interaction
for which the N\'{e}el temperature takes a maximum value.
\cite{ref-Rev1},\cite{ref-Ulmke}.
Since the strength of the interactions we studied 
is $U \simeq U_{c}$ \cite{ref-3D},
we did not observe it.

In summary,
the three-dimensional Hubbard model with random potential 
of a finite range has been studied numerically
using a finite temperature quantum Monte Carlo method.
The temperature dependence of the charge compressibility
suggests that sufficiently strong disorder closes the Mott gap.
The transition from an incompressible phase to a compressible phase
occurs at a finite strength of disorder.
The disorder also destroys an antiferromagnetic long-range order
which is characteristic of the Mott insulator.
As in the case of the Mott gap,
the antiferromagnetic correlation is robust against weak disorder.
These features are common
in one, two and infinite dimensional systems.

We are grateful to 
Morita, 
Kawakami, 
Fujimoto, 
Denteneer,
R\"omer and 
Kishine
for helpful correspondence.
Y.H is supported in part by  Grant-in-Aid
from the Ministry of Education, Science and Culture of Japan.
The computation in this work has been done
using the facilities of the Supercomputer Center,
ISSP, University of Tokyo.


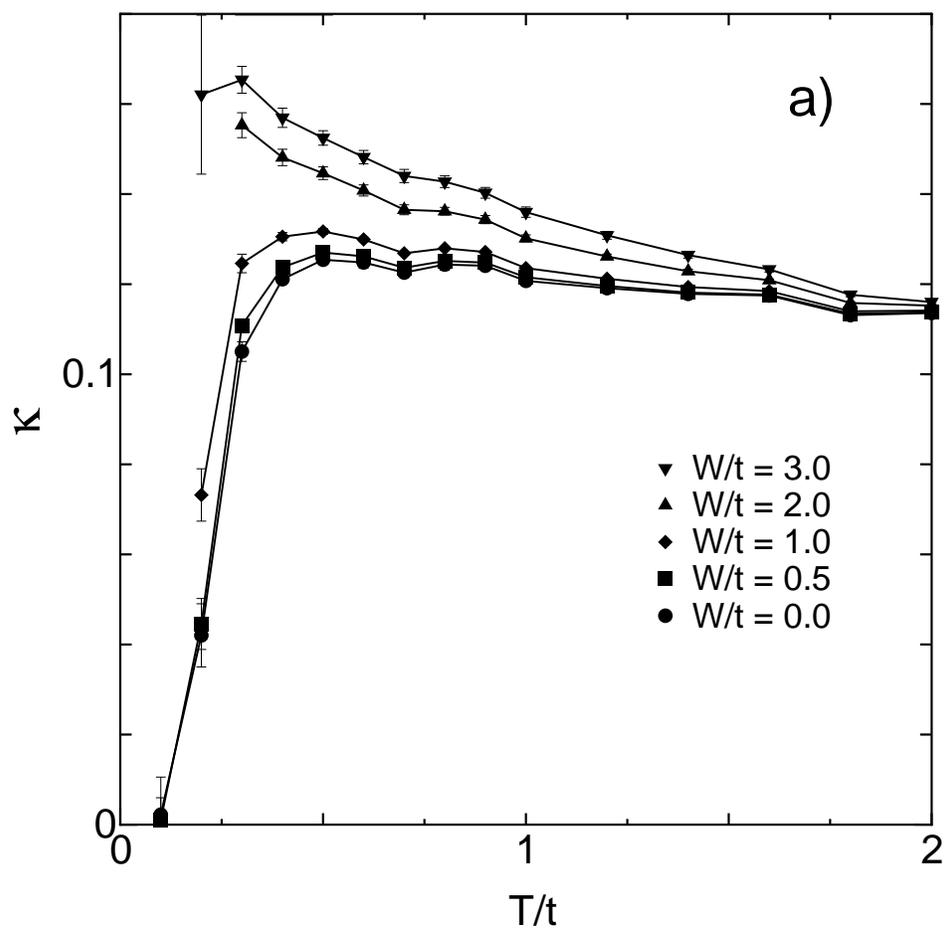
\begin{figure}
\caption{
Temperature dependence of 
the charge compressibility $\kappa$,
where $U/t=6$, a) $L = 4 \times 4 \times 4$ and b) $L = 6 \times 6 \times 6$.
Without disorder,
$\kappa$ shows 
thermally-activated behavior 
and decreases 
toward $T=0$,
indicating the existence of a charge gap.
For weak disorder,
$\kappa$ still shows thermally-activated behavior. 
On the other hand, for strong disorder,
$\kappa$ does not decrease down to the temperature we studied.
It is a disorder-driven transition from an incompressible phase 
to a compressible phase.}
\end{figure}

\begin{figure}
\caption{
The antiferromagnetic structure factor as a function of temperature ($T/t$),
where $U/t=6$, a) $L = 4 \times 4 \times 4 $ and b) $L = 6 \times 6 \times 6$.
For weak disorder,
the antiferromagnetic structure factor shows diverging behavior
down to the temperature we studied.
On the other hand, for strong disorder,
the divergence behavior is not observed.
This indicates that sufficiently strong disorder destroys
the long-range antiferromagnetic correlation 
which is characteristic of the Mott insulator.}
\end{figure}

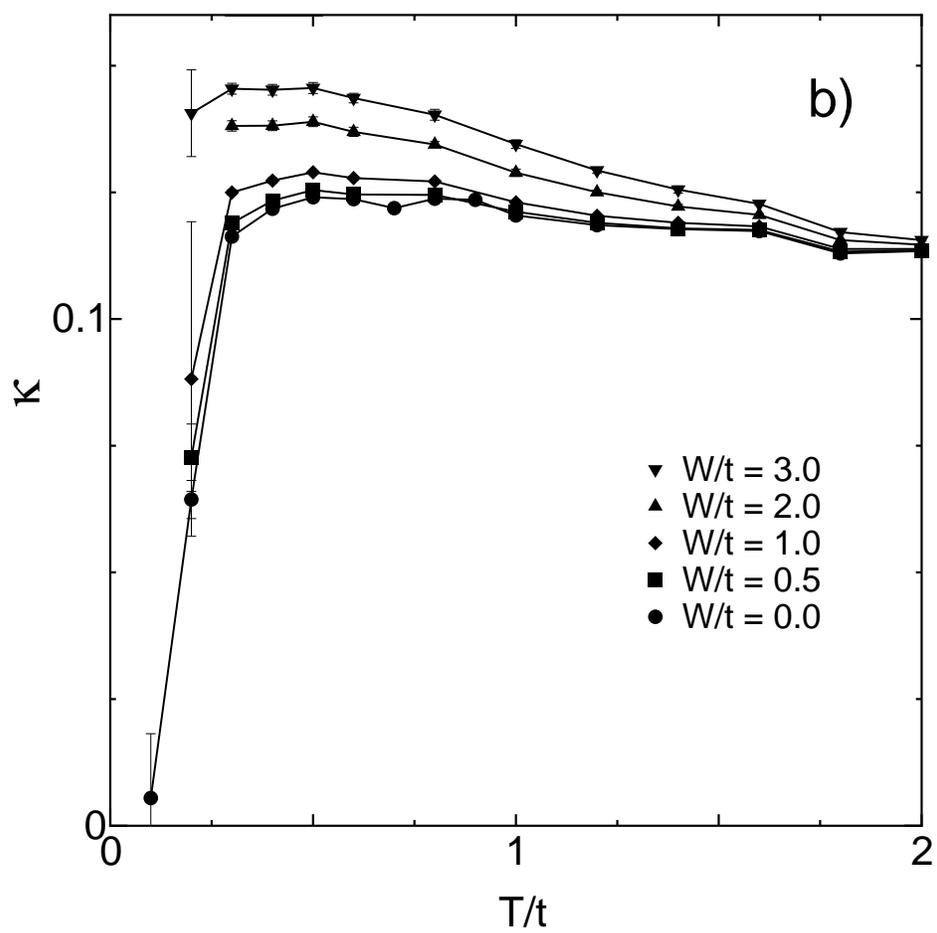
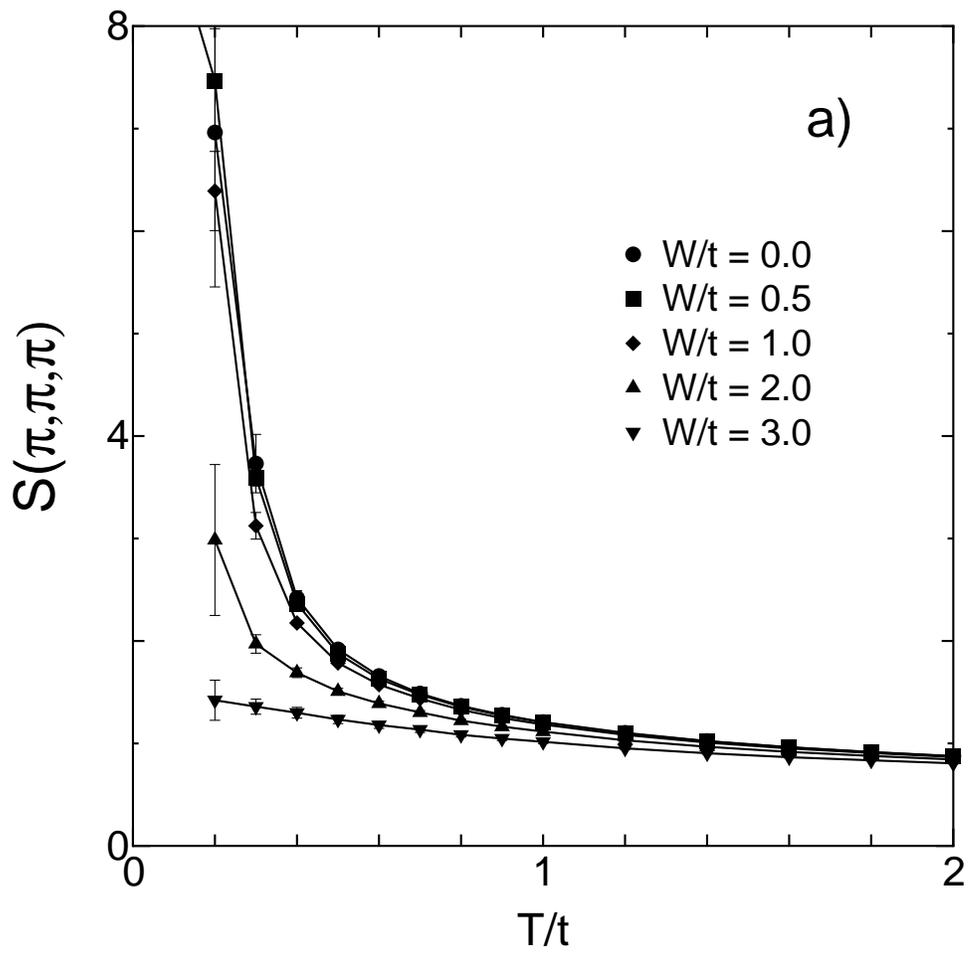
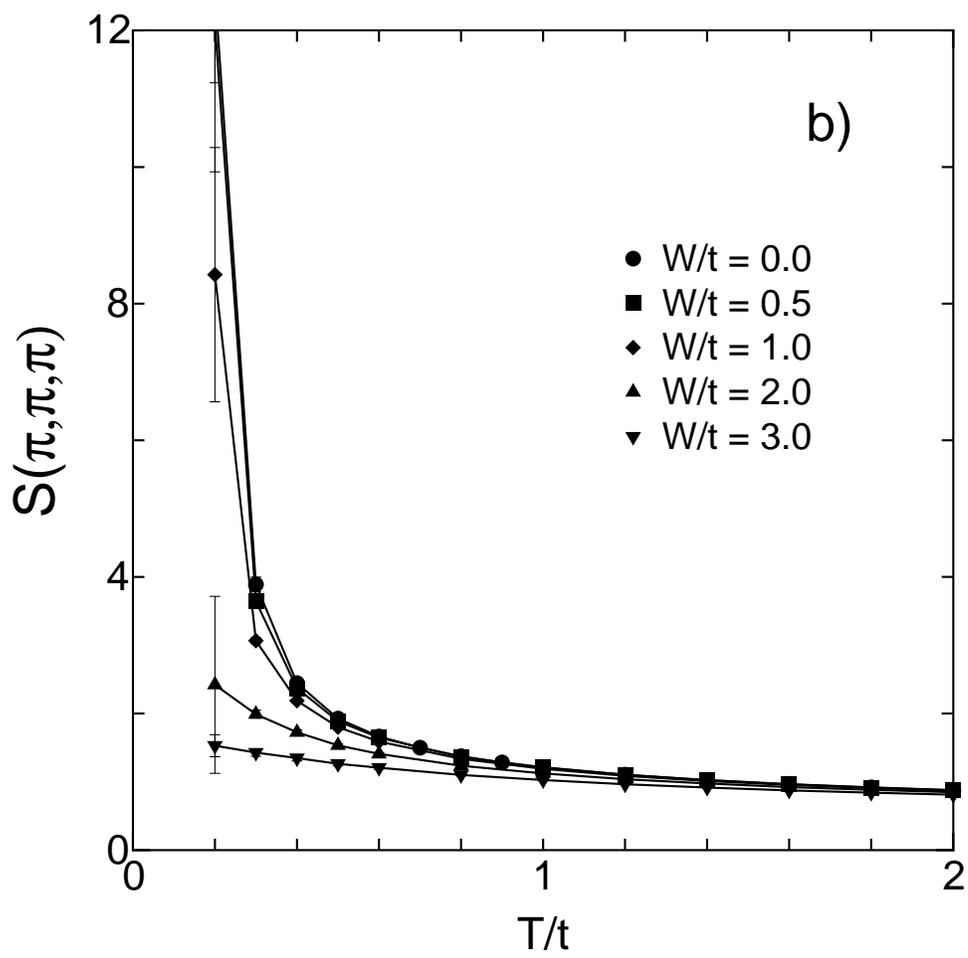
\end{document}